\documentclass[galley,jgrga]{agu2001}

\usepackage{lineno}
\usepackage{graphicx}
\authorrunninghead{SARTELET ET AL.}

\titlerunninghead{Aerosol processes over Greater Tokyo}

\authoraddr{K. N. Sartelet,
CEREA, Ecole Nationale des Ponts et Chauss\'ees,
  6-8 Avenue Blaise Pascal, 77455 Champs sur Marne, France.
(sartelet@cerea.enpc.fr)}

\begin{document}

\title{Dominant aerosol processes during high-pollution episodes over Greater Tokyo}

\author{K.N. Sartelet}
\affil{CEREA, Research Center for Atmospheric Environment, Joint
  Laboratory Ecole Nationale des Ponts et Chauss\'ees and Electricit\'e
  de France, France}

\author{H. Hayami}
\affil{Central Research Institute of Electric Power Industry, Abiko,
Chiba, Japan}

\author{B. Sportisse}
\affil{CEREA, Research Center for Atmospheric Environment, Joint
  Laboratory Ecole Nationale des Ponts et Chauss\'ees and Electricit\'e
  de France, France}
\begin{abstract}

This paper studies two high-pollution episodes over Greater Tokyo: 9 and 10 December 1999, and 31 July and 1 August 2001. 
Results obtained with the chemistry-transport model (CTM) Polair3D are compared to measurements of inorganic PM$_{2.5}$. 
To understand to which extent the aerosol processes modeled in Polair3D impact simulated inorganic PM$_{2.5}$, Polair3D is run with different options in the aerosol module, e.g. with/without heterogeneous reactions. To quantify the impact of processes outside the aerosol module, simulations are also done with another CTM (CMAQ). 

In the winter episode, sulfate is mostly impacted by condensation, coagulation, long-range transport, and deposition to a lesser extent. In the summer episode, the effect of long-range transport largely dominates. 
The impact of condensation/evaporation is dominant for ammonium, nitrate and chloride in both episodes. However, the impact of the thermodynamic equilibrium assumption is limited. The impact of heterogeneous reactions is large for nitrate and ammonium, and taking heterogeneous reactions into account appears to be crucial in predicting the peaks of nitrate and ammonium.
The impact of deposition is the same for all inorganic PM$_{2.5}$. It is small compared to the impact of other processes although it is not negligible.
The impact of nucleation is negligible in the summer episode, and small in the winter episode. The impact of coagulation is larger in the winter episode than in the summer episode, because the number of small particles is higher in the winter episode as a consequence of nucleation.

\end{abstract}

\begin{article}

\section{Introduction}

With the impact of air pollution on health and vegetation being a great concern, chemical transport models (CTMs) are often used at a regional scale to predict air quality, i.e. to compute the distribution of atmospheric gases, aqueous phase species and particulate matter. High pollution episodes may occur depending on emissions and sources of pollutants, as well as on the dynamical characteristics of the meteorology. 

The meteorology over Greater Tokyo is strongly influenced by orography and sea/land breeze regimes, as shown by \cite{fujibe} and \cite{ohara}. High pollution episodes are often observed in early summer and early winter. Early winter episodes are often a consequence of the meso-synoptic scale meteorology (\cite{mizuno}, \cite{uno}, \cite{ohara2}): sea/land breeze and the blocking effect of orography in central Japan. These episodes are characterized by high NO$_2$ and chloride concentrations (\cite{uno}, \cite{kaneyasu}). Measurements made by \cite{kaneyasu} suggest that the precursor HNO$_3$ of aerosol nitrate is formed by photochemical reactions, while the precursor of aerosol chloride may be locally emitted. For summer episodes,
the role of sea breeze penetration for oxidants is detailed by \cite{wakamatsu}. \cite{hiroshi} founds that the PM mass completely changes in proportion to RH in an episode that happened within 24 hours. Rapidly increasing RH may enhance condensation onto aerosols.

Two episodes are studied in this paper: one early winter episode (9 and 10 December 1999), and one summer episode (31 July and 1 August 2001). These episodes are simulated using the CTM Polair3D (\cite{bs-ijep}). 
This paper aims at studying the impact of different aerosol processes that may influence the particle concentrations observed during these episodes. Furthermore, the impact of the aerosol processes is compared to the impact of numerical choices made in the aerosol model. For example, the impact of the size distribution is studied by replacing the modal aerosol model MAM (\cite{sartelet-mam}) used in Polair3D by a sectional model (SIREAM: \cite{debry06siream}). Aerosol concentrations are not only influenced by the aerosol model, but also by the parameterizations and the numerical schemes used for advection, diffusion, chemical mechanism (\cite{mallet05uncertainty}). To have an estimate of how these processes may influence the concentrations, the impact of the different aerosol processes is compared to the impact of using a different CTM (CMAQ, \cite{bink-cmaq}, \cite{eder}, \cite{cmaqo3}, \cite{eta-cmaq}).

\section{The Models}

The CTM Polair3D (\cite{bs-ijep}) is used with the chemical mechanism RACM (\cite{racm}). Photolysis rates are computed off-line, as
done in the photolysis rate preprocessor of CMAQ (\cite{cmaq14}). Vertical diffusion is computed using the Troen and Mahrt's parameterization (\cite{troen}) within the boundary layer, and using the Louis' parameterization (\cite{louis}) above it. 
Polair3D may be used with two aerosol models : MAM (\cite{sartelet-mam}) and SIREAM (\cite{debry06siream}). The difference between the models lies in the size distribution: in MAM the size distribution is modeled with four log-normal modes (modes $i$, $j$, $k$, $c$), and in SIREAM it is modeled with sections. Four sections are used in the simulations of this paper. In MAM and SIREAM, the modes/sections are bounded as follow: mode~$i$: $< 0.01 \mu$m, mode~$j$: [$0.01; \; 0.1 \mu$m], mode~$k$: [$0.1; \; 2.5 \mu$m], mode~$c$: $> 2.5 \mu$m.
A complete technical description of MAM and SIREAM may be found in \cite
{sportisse06siream}. MAM is used for the simulations of this paper, except when specified. It is now briefly described.

\subsection{Composition}

Particles can be made of inert species (dust and elemental carbon), liquid water, inorganic species (sodium, chloride, ammonium, nitrate and sulfate) and organic species.

\subsection{Aerosol Processes}

 Coagulation, condensation/evaporation and nucleation are modeled as described in \cite{sartelet-mam}. 

\subsubsection{Condensation/evaporation}
Condensation/evaporation is computed using the thermodynamic module
ISORROPIA (\cite{ed_isorropia}). By default, thermodynamic equilibrium is assumed
between the gas and the aerosol phases. However, Polair3D may also
be used without the assumption of thermodynamic equilibrium for large
modes/bins, for which condensation/evaporation is then computed
dynamically. Although the thermodynamic equilibrium assumption may not
be valid especially for large aerosols, it is used in the baseline simulations here because it is
computationally efficient. The effect of this assumption is studied in a separate sensitivity simulation

\subsubsection{Nucleation}
Nucleation is modeled using the ternary parameterization (water,
ammonium and sulfate) of Napari (\cite{napari}). The ternary
nucleation rate is several order of magnitudes larger than commonly
used binary nucleation rates (water and sulfate).

\subsubsection{Heterogeneous reactions}
In the first runs presented here, heterogeneous reactions are not taken into account in Polair3D. If heterogeneous reactions are taken into account in Polair3D, they are modeled according to \cite{jacob2000}: HO$_2$ $\rightarrow$ 0.5 H$_2$O$_2$, NO$_2$ $\rightarrow$ 0.5 HONO + HNO$_3$, NO$_3$ $\rightarrow$ HNO$_3$, N$_2$O$_5$ $\rightarrow$ 2 HNO$_3$.
The kinetic rates of these first order reactions are $k_i=\left(\frac{a}{D_i^g}+\frac{4}{\bar{c}_i^g\gamma}\right)^{-1}S_a$, where
$a$ is the particle radius, $\bar{c}_i^g$ is the gas-phase thermal velocity in
the air, $S_a$ is the available surface of condensed matter per air
volume, and $\gamma$ is the reaction probability that a molecule impacting the aerosol surface undergoes reaction. $\gamma$ strongly depends on the chemical and size distribution of
particles. The values used in this paper are $\gamma_{HO_2} = 0.2$, 
$\gamma_{NO_2} =  10^{-4}$, $\gamma_{NO_3}=10^{-3}$ and
$\gamma_{N_2O_5} =  0.03$. 

\subsubsection{Dry deposition}

Dry deposition is parameterized with a resistance approach following \cite{depaerosol}. The processes modeled include gravitational settling and the deposition processes of Brownian diffusion, impaction, interception, particle rebound. The aerodynamic resistance and the friction velocity are computed as in CMAQ (\cite{bink-cmaq}). Dry deposition depends on the diameter of the particles. In Polair3D-MAM, for each log-normal mode and each moment, the deposition velocity is integrated over diameters by fourth order Gauss-Hermite quadratures. In Polair3D-SIREAM, the mean diameter of each section is used to compute the deposition velocities.

\subsubsection{Mode Merging versus Mode splitting}

A mode merging scheme or a mode splitting scheme is required in a modal model to prevent modes from overlapping, i.e. to force modes to be of distinct size ranges throughout the simulations. 
Different mode merging schemes may be used, often based on that of \cite{bink-cmaq}, where the threshold diameter between the two modes to be merged is chosen as the diameter where the number distributions of the two modes overlap. In Polair3D, mode merging is applied between modes $i$ and $j$ (and between modes $j$ and $k$), when the diameter of the volume distribution of mode $i$ (and mode $k$) exceeds a fixed diameter of $0.01\mu m$ for mode $i$ (and of $0.1\mu m$ for mode $j$). 

In the tests performed later in this paper, the mode merging scheme is replaced by a mode splitting scheme (\cite{sartelet-mam}), which is designed to reproduce the evaluation of a mode that splits into two modes under the combined effect of nucleation, condensation and coagulation.

For the simulations done with CMAQ (\cite{bink-cmaq}, \cite{mebust-cmaq}, \cite{cmaq}), the version 4.3 is used, with the Carbon Bond IV chemical mechanism (\cite{cbm4}). It is modified to include sodium and chloride in the Aitken and in the accumulation modes.
In CMAQ, the aerosol module is a modal one with 3 modes. As in Polair3D, the following processes are taken into account: coagulation, condensation/evaporation, nucleation, heterogeneous reactions, dry deposition. For condensation/evaporation, thermodynamic equilibrium is assumed between the gas and the aerosol phases. Nucleation is modeled with a binary nucleation rate. The heterogeneous reaction N$_2$O$_5$ $\rightarrow$ 2 HNO$_3$ is taken into account.

\section{Domain and Input Data}

\begin{figure}
	\begin{center}
		\includegraphics[angle=0,width=0.75\textwidth]{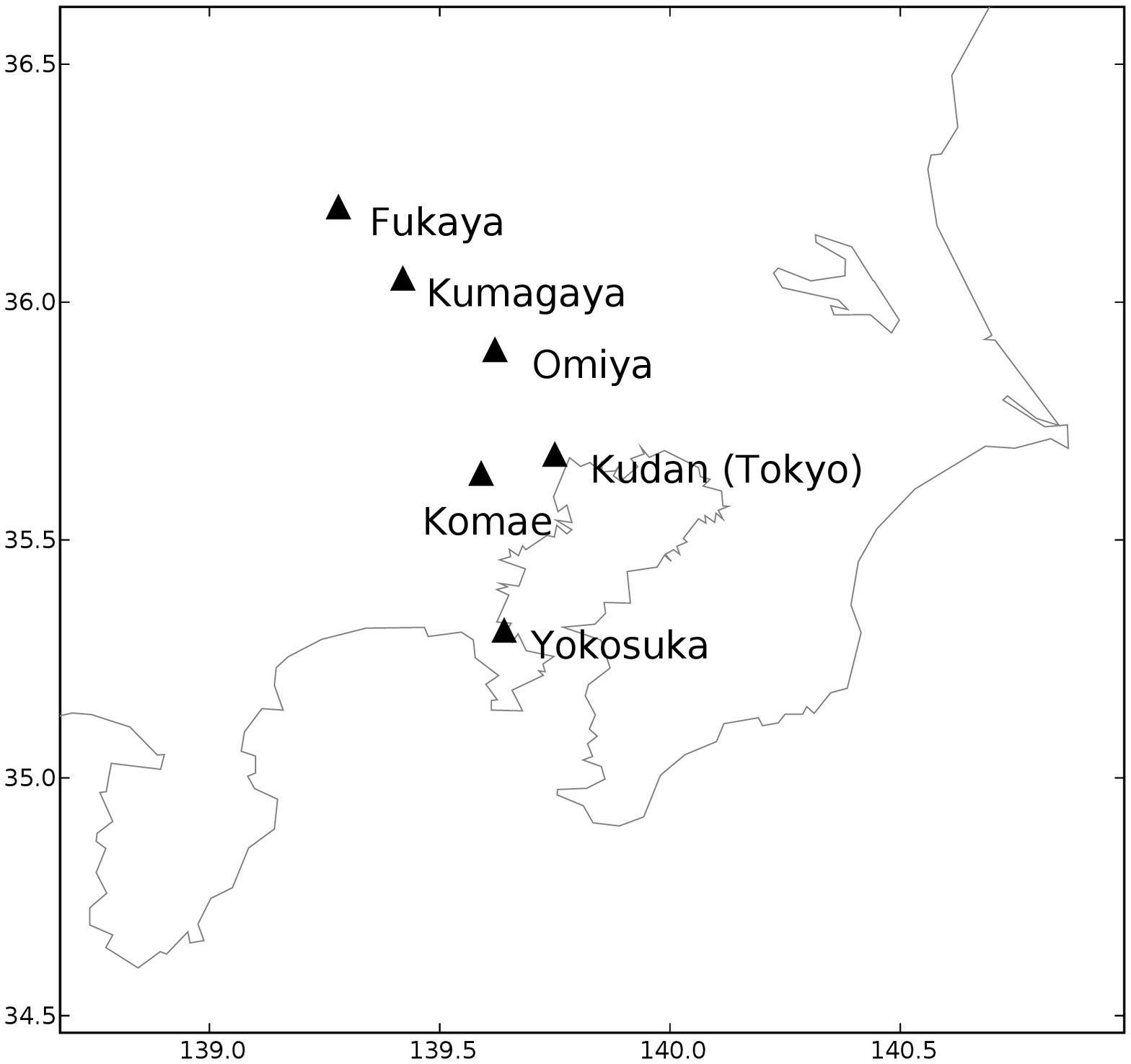}
	\end{center}
\caption{Location of stations at which the comparisons to data are made.}
\label{stations}
\end{figure}

Simulations are performed over a 210km x 240km area, centered around
Tokyo, with a 5km x 5km resolution (Figure~\ref{stations} shows the domain of simulation, which is discretized with 42 x 48 points). The horizontal domain is the same for the simulations done with Polair3D and CMAQ.
For simulations with Polair3D, 12 vertical layers are
considered (0m, 29m, 58m, 103m, 147m, 296m, 447m, 677m, 954m, 1282m, 1705m, 2193m, 2761m). A no-flux boundary condition (free atmosphere) is used at the top boundary for diffusion. In CMAQ, 16 layers of varying thickness extend to about 16km height. The vertical coordinate is not altitude but sigma-levels. The averaged altitudes of the first 10 layers of CMAQ correspond to the altitudes used in Polair3D. 
Meteorological data are provided by the Japanese Meteorological
Agency with a 20km x 20km resolution every six hours. Finer hourly
meteorological data, with a 5km x 5km resolution are obtained by
running the meteorological model MM5, the Fifth-Generation Pennsylvania State University/National Center for Atmospheric Research (NCAR) Mesoscale model (\cite{mm5}). Initial and boundary conditions
(with inputs varying every three hours) are obtained by running the
CTM CMAQ over East Asia with a 45km x 45km resolution. Emission
inventories are provided by a collaboration with the Japanese 
National Institute for Environmental Studies (\cite{HH-em}). Emission sources
include mobile sources (road, air, vessels), stationary sources
(domestic, industries), waste water treatment, biogenic/natural
sources (agriculture, soil, volcanoes). Table~\ref{emtot} summarizes the total amount of emitted particulate matter and precursors over the domain. The emission inventory does not contain information about either the size distribution or the chemical speciation. The size distribution and the chemical speciation of $PM_{10}$ and $PM_{2.5}$ are specified as in CMAQ (\cite{bink-cmaq}). All $PM_{10}-PM_{2.5}$ are assigned to the coarse mode and particles are assumed to be made of $90\%$ dust and $10\%$ elementary carbon. Most of $PM_{2.5}$ ($99.9\%$) are assigned to the accumulation mode, and $0.1\%$ to the Aitken mode. For $PM_{2.5}$, particles are assumed to be made of $30\%$ dust and $70\%$ elementary carbon. The parameters of the three modes used for emission are shown in Table~\ref{emsize}. Whereas the modal model in CMAQ has three modes (aitken, accumulation and coarse), the modal model in MAM has an additional mode designed for nanometer nucleated particles. This mode is not used for emission.

\begin{table}[htbp]
  \caption{Total amount (in kt) of emitted particulate matter and precursors over the domain (210km x 240km) for each episode.}
\begin{flushleft}
  \begin{tabular}{lcccccc}
\tableline
& $PM_{2.5}$ & $PM_{10}$ &  $NO_x$ & $SO_x$ & $NH_3$ & $HCl$ \\
\tableline
Winter episode & 0.44 & 0.23 & 3.72 & 1.75 & 0.54 & 0.16 \\
Summer episode & 0.40 & 0.23 & 2.82 & 1.24 & 1.81 & 0.15 \\
\tableline
  \end{tabular}
\end{flushleft}
  \label{emtot}
\end{table}

\begin{table}[htbp]
  \caption{Parameters of the three modes used for emission.}
\begin{flushleft}
  \begin{tabular}{lcccccc}
\tableline
& Aitken & Accumulation &  Coarse \\
\tableline
Mean Diameter ($\mu$m) & 0.03 & 0.3 & 6 \\
Standard Deviation & 1.7 & 2 & 2.2 \\
\tableline
  \end{tabular}
\end{flushleft}
  \label{emsize}
\end{table}

For gaseous species, dry deposition velocities are computed off-line following \cite{wesely}.

Simulations start one day before the episode to allow for spin-up. Tests using two days for spin-up showed that one day spin-up is sufficient for the gaseous and $PM_{2.5}$ concentrations considered here.

\section{Description of The Episodes}

\subsection{9$^{th}$ and 10$^{th}$ December 1999}

\begin{figure}
	\begin{center}
		\includegraphics[angle=-90,width=0.45\textwidth]{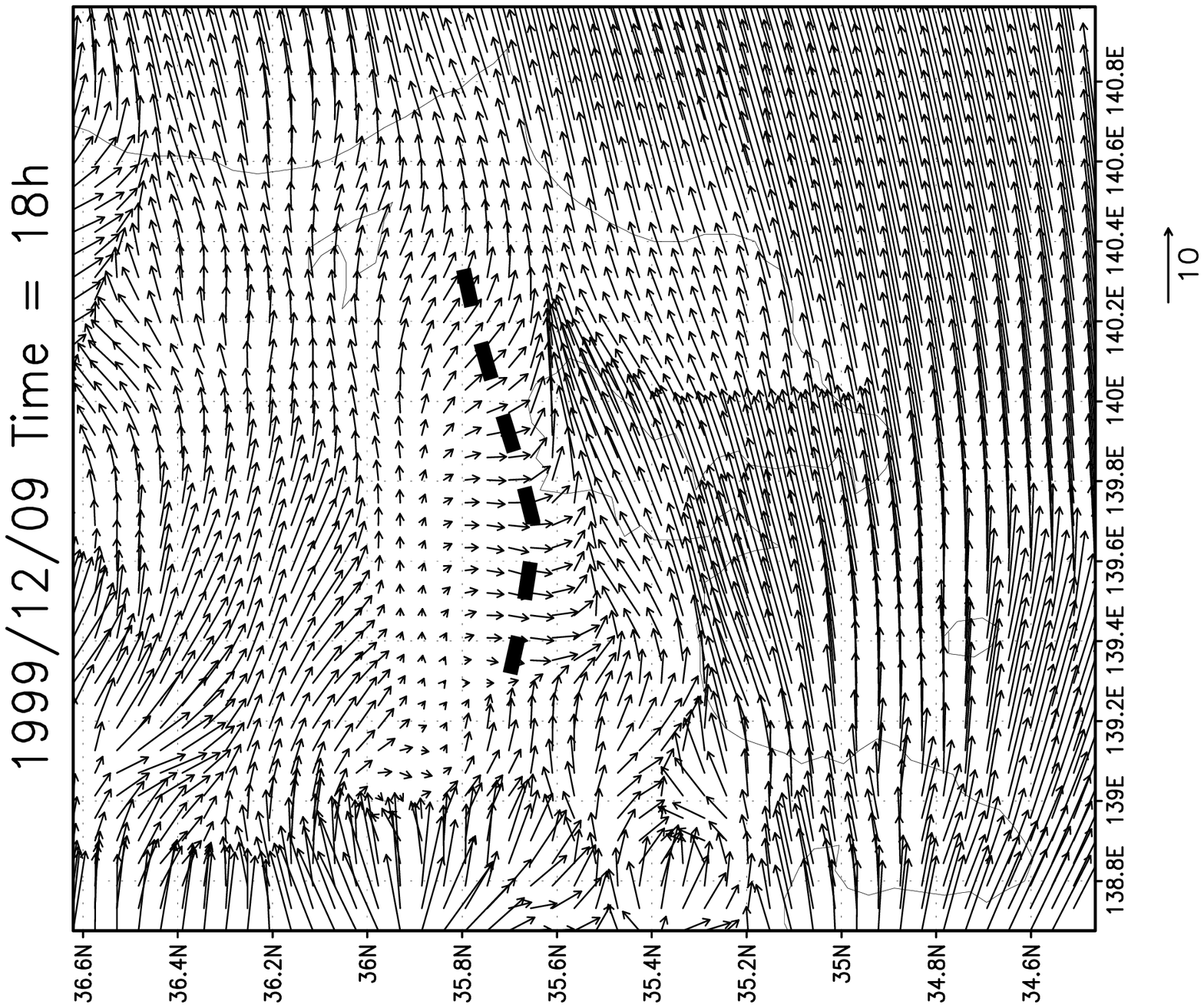}
	\end{center}
\caption{Wind vectors on 9 December at 6pm at 14.5m height (MM5 results). Black dashed line: observed meso-front.}
\label{fig:9-10meteo}
\end{figure}

\begin{figure}
	\begin{center}
		\includegraphics[angle=-90,width=0.90\textwidth]{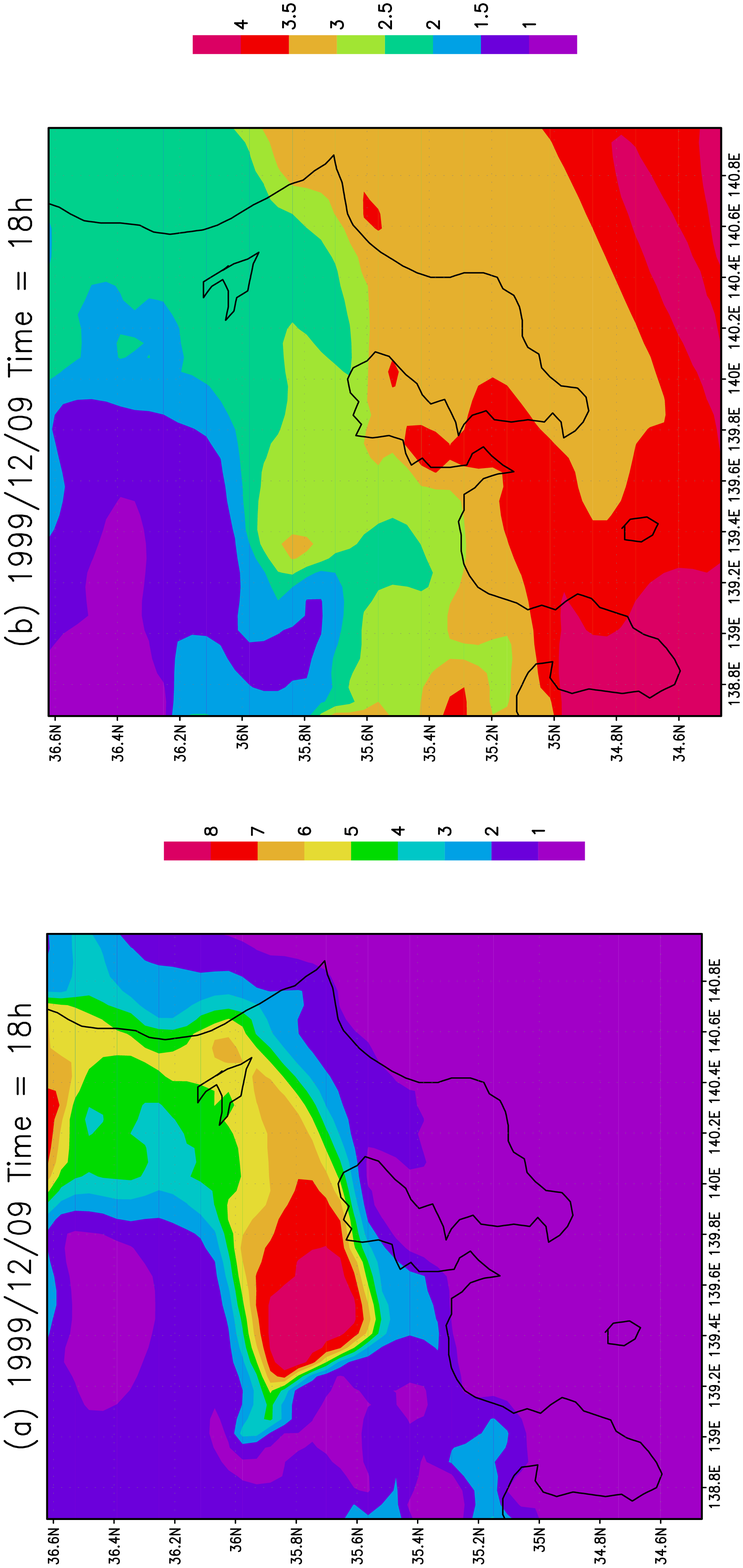}
	\end{center}
\caption{Modeled nitrate (panel a) and sulfate (panel b) concentrations on 9 December at 6pm.}
\label{fig:9conc}
\end{figure}

During this episode, chemical concentrations are observed to be high. On the 9$^{th}$, this is mostly due, to the presence of a meso-front. 
Strong cold winds coming from West and North West stay weak at low altitudes because of the presence of orography in the West and in the North West. Warm winds from South do not penetrate much in land and a meso-front is observed (black dashed line in Figure~\ref{fig:9-10meteo}).  As shown in Figure~\ref{fig:9-10meteo}, 
the meso-front is not so well reproduced by MM5. 
On the 9$^{th}$, pollutants accumulate in the Northern part of the front. This is seen in Figure~\ref{fig:9conc}, which shows high nitrate concentrations in the Northern part of the front at 6pm. Sulfate concentrations are also high where ammonium and nitrate concentrations are high, but sulfate concentrations are even higher over sea. These high concentrations are brought from South West, but they do not propagate much over land because of the meso front.
On the 10$^{th}$, high concentrations are a consequence of weak winds and high pollution. 

\subsection{31$^{st}$ July and 1$^{st}$ August 2001}

The episode of 31 July and 1 August 2001 is characteristic of sea and land breeze circulation. Wind patterns are also influenced by local winds.

\begin{figure}
	\begin{center}
		\includegraphics[angle=-90,width=0.37\textwidth]{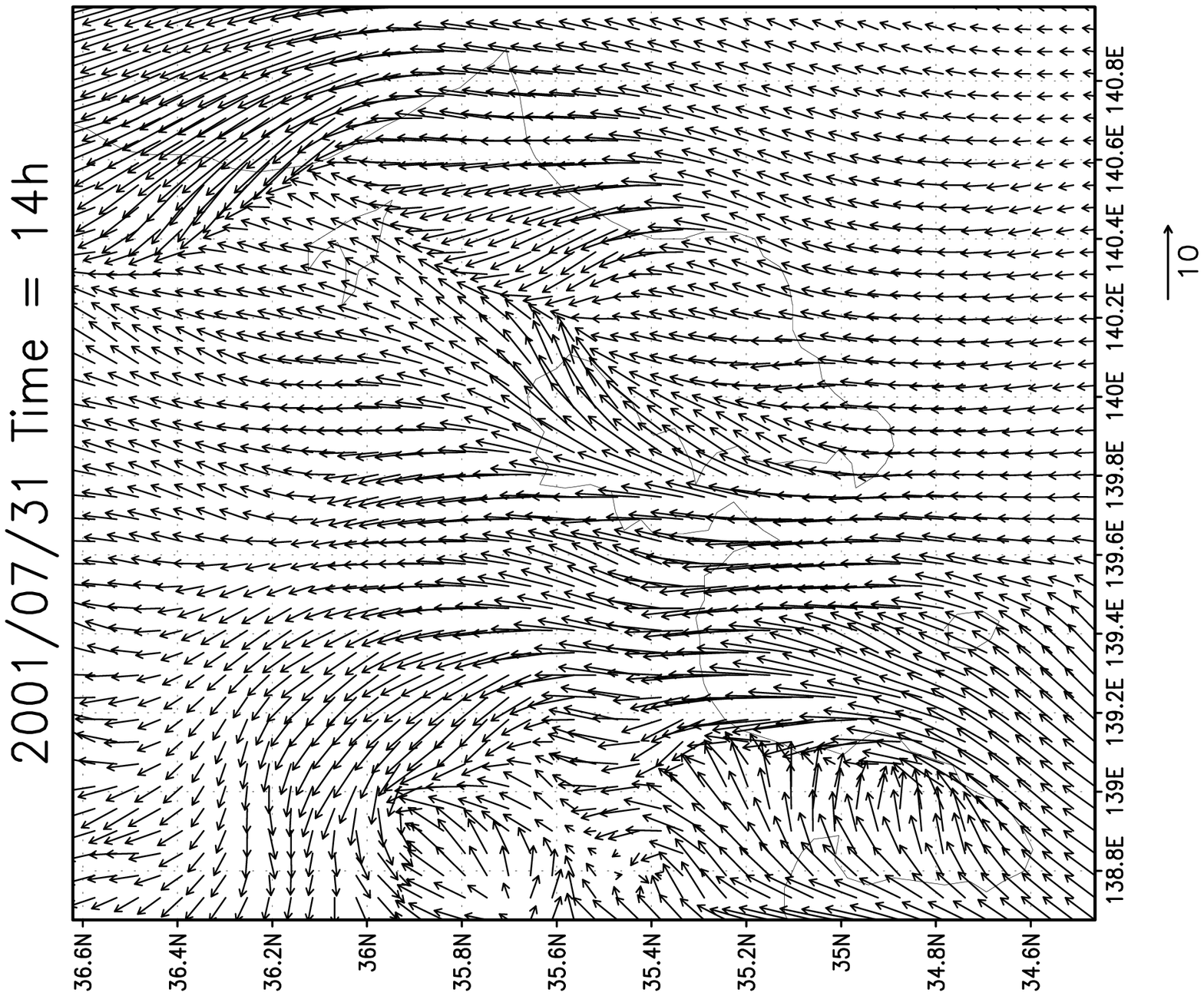}
	\end{center}
\caption{Wind vector at 2pm on 31 July.}
\label{fig:31meteo}
\end{figure}

\begin{figure}
	\begin{center}
		\includegraphics[angle=-90,width=0.45\textwidth]{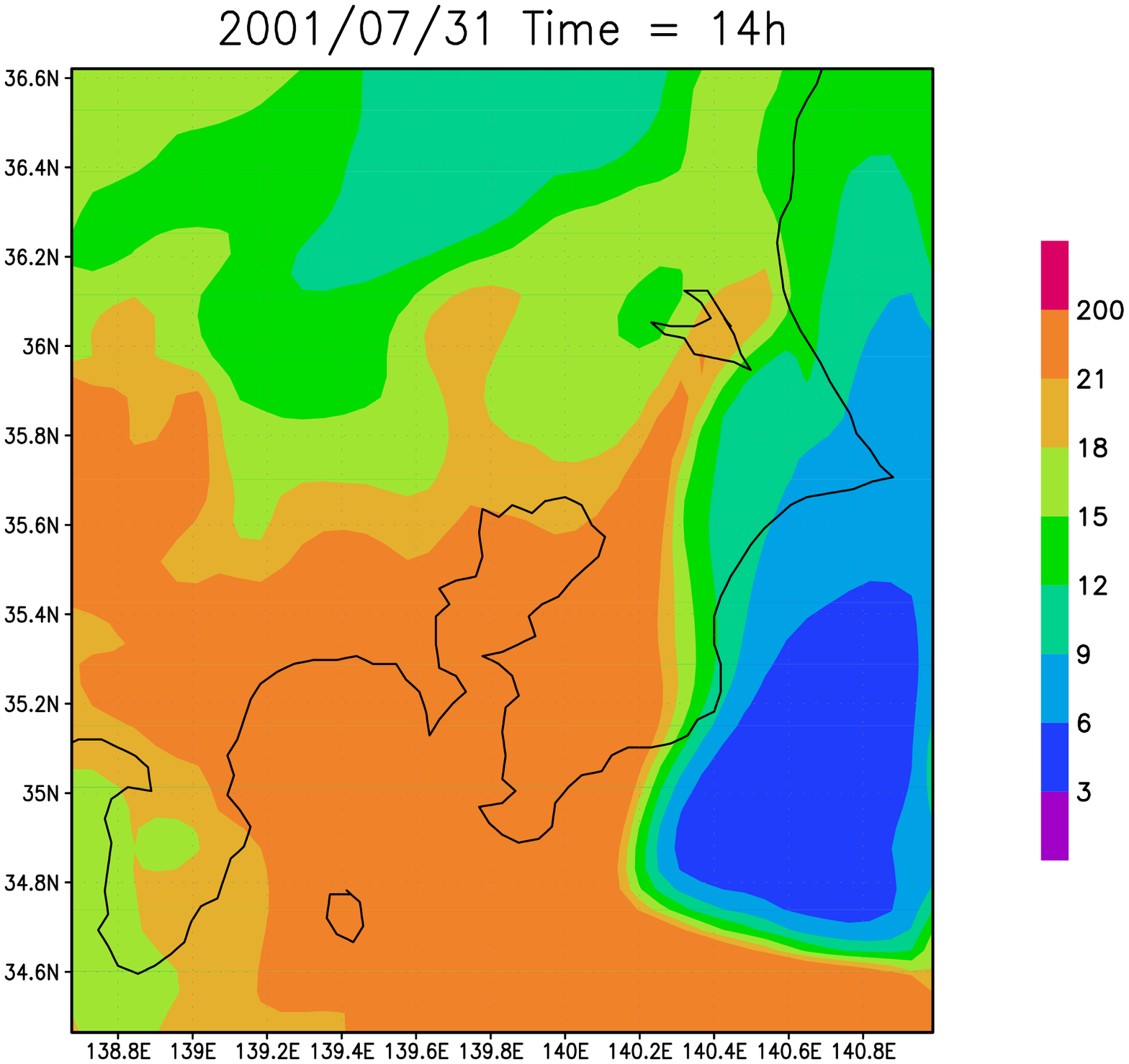}
	\end{center}
\caption{Modeled sulfate concentrations on 31 July at 2pm.}
\label{fig:31conc}
\end{figure}

On 31 July, winds are weak over the whole domain. Winds from west and south west strengthen during the day, and penetrate on land as the temperature on land gets warmer than the temperature on sea. Figure~\ref{fig:31meteo} displays the wind vector at 2pm on 31 July. High sulfate concentrations are brought from the south west. At 2pm, sulfate concentrations are high over almost the whole domain, as can be seen in Figure~\ref{fig:31conc}.

On 31 July at night, although south westerly winds are strong over sea, winds stay low on land. In the early morning of 1 August, south westerly winds weaken. They do not penetrate much on land, and winds on land stay low.

\section{Comparison to Measurements}

\subsection{Measurements}

Aerosol distributions were collected using impactor, denuder, and filter pack (\cite{hayami-meas}), every three hours, except for the summer 2001 where at night they were collected every six hours.
Hourly gaseous distributions at
different stations distributed over Greater Tokyo are provided by
Japanese local institute governments. 

The stations at which comparisons are made are shown in Figure~\ref{stations}.

On 9 and 10 December 1999, for gas, data are available for O$_3$, NO$_x$ and SO$_2$ at 3 sites: Yokosuka, Fukaya and Kudan. For PM$_{2.5}$, data are available for sulfate, ammonium, nitrate and chloride at 4 sites: Yokosuka, Omiya, Fukaya and Kudan (Tokyo). 

On 31 July and 1 August 2001, data are available at two sites: Kumagaya and Komae, for O$_3$, NO$_x$, SO$_2$ and PM$_{2.5}$.

Sulfate concentrations are overall higher on 31 July and 1 August 2001 than on 9 and 10 December 1999, with an average over the stations of 13.3$\mu$g~m$^{-3}$ against 2.6$\mu$g~m$^{-3}$. Furthermore, sulfate represents as much as 50$\%$ of the inorganic PM$_{2.5}$ in the summer episode, against 21$\%$ in the winter one. Ammonium and nitrate concentrations are also higher on 31 July and 1 August 2001 than on 9 and 10 December 1999, with an average of 7.7$\mu$g~m$^{-3}$  against 2.7$\mu$g~m$^{-3}$  for ammonium, and 5.7$\mu$g~m$^{-3}$  against 4.5$\mu$g~m$^{-3}$  for nitrate.

\subsection{Statistics}

Comparison of the results obtained with Polair3D to measurements is done for inorganic fine aerosols (PM$_{2.5}$). Model results for this comparison take into account data only from the grid boxes for which observations are available. In MAM/SIREAM, PM$_{2.5}$ are computed by summing the three smallest modes/sections (modes~$i$, $j$ and~$k$). Similarly, in CMAQ, as the mode~$i$ is not modeled, the two smallest modes, i.e.~$j$ and~$k$, are summed up to compute PM$_{2.5}$

The relative bias and error between modeled and observed concentrations are quantified using unbiased symmetric metrics (\cite{yu-metrics}): the normalized mean bias factor B$_{NMBF}$ and the normalized mean absolute error factor E$_{NMAEF}$ (Appendix A). 
%The mean fractional bias MFB and the mean fractional error MFE defined by \cite{boylan} are shown for comparison to the performance criteria defined by \cite{boylan}.
The correlation coefficient ($\%$) is also used as a statistical indicator. The smaller the bias and the error are, and the larger the correlation is, the closer the model fits the observation. Bias indicates whether the model tends to under or over-predict the observation, and error indicates how large the deviation is. 

\cite{yu-metrics} suggests the criteria of model performance for sulfate to be taken as $\left| B_{NMBF} \right| \leq 25\%$ and $E_{NMAEF} \leq 35\%$.

\subsection{9$^{th}$ and 10$^{th}$ December 1999}

The statistical indicators are shown in Table~\ref{RMSE1999}.  The model satisfies the criteria suggested by \cite{yu-metrics} for sulfate. However, these criteria are not satisfied for ammonium, nitrate and chloride, which are more difficult to model because of their volatility.
This difficulty to model nitrate and ammonium is stressed for example by \cite{zhang-mm5} where nitrate and ammonium are underpredicted by factors 9.6 and 2.1 respectively in the Southeastern US for the period of 1-10 July 1999. 

\begin{figure}
	\begin{center}
		\includegraphics[angle=0,width=0.95\textwidth]{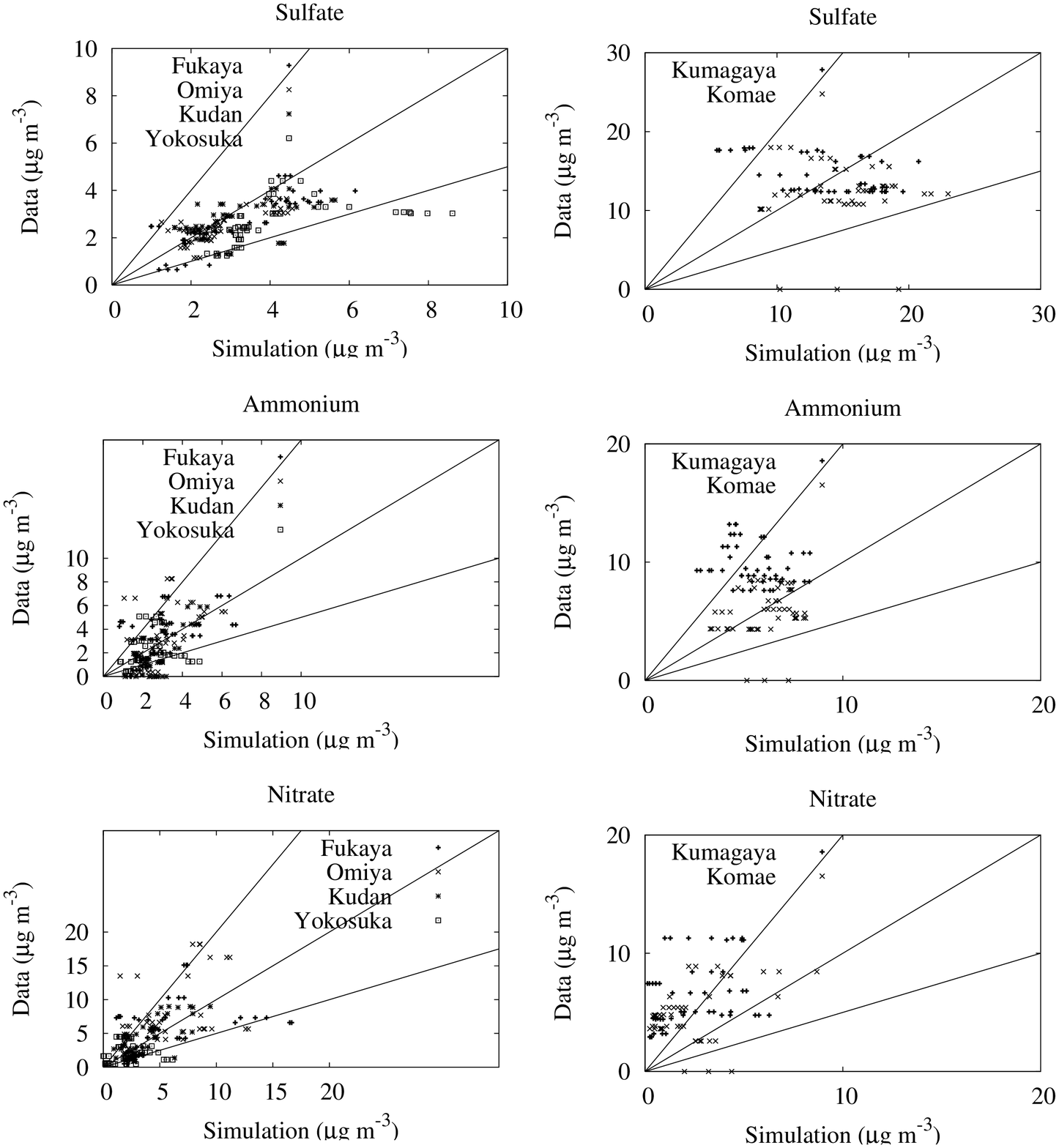}
        \end{center}
\caption{Scatter plots of observation (ordinate) versus simulation (abscissa) for sulfate, ammonium and nitrate for 9 and 10 December 1999 (left panels) and for 31 July and 1 August 2001 (right panels). 1:1, 1:2, and 2:1 reference lines are provided.}
\label{sp}
\end{figure}

The correlation coefficients are good for all species, ranging from $66\%$ for sulfate to $36\%$ for chloride. This suggests that the overall diurnal patterns are well modeled. 
As shown in Figure~\ref{sp}, the majority of hourly simulations falls within a factor 2 of the observations for sulfate and for high values of nitrate and ammonium. However, values lower than about 3$\mu$g~m$^{-3}$  for ammonium and 5$\mu$g~m$^{-3}$ for nitrate are often scattered outside the factor of two reference line.

\begin{figure}
	\begin{center}
	  \includegraphics[width=0.95\textwidth]{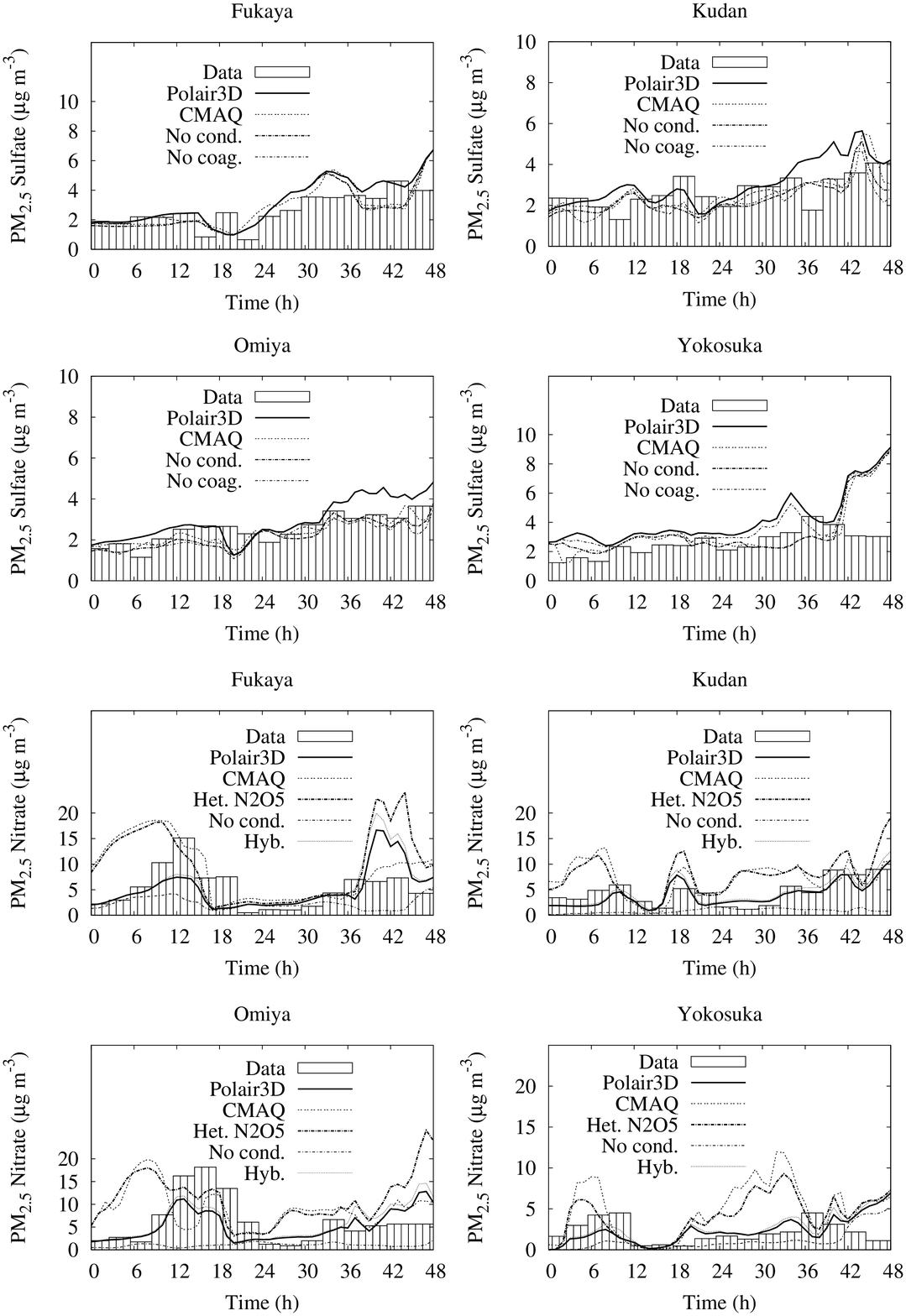}
       \end{center}
\caption{$PM_{2.5}$ concentration of sulfate and nitrate at Fukaya, Yokosuka, Kudan, Omiya for 9 and 10 December 1999.}
\label{pm1999}
\end{figure}

\begin{figure}
	\begin{center} 
	  \includegraphics[width=0.95\textwidth]{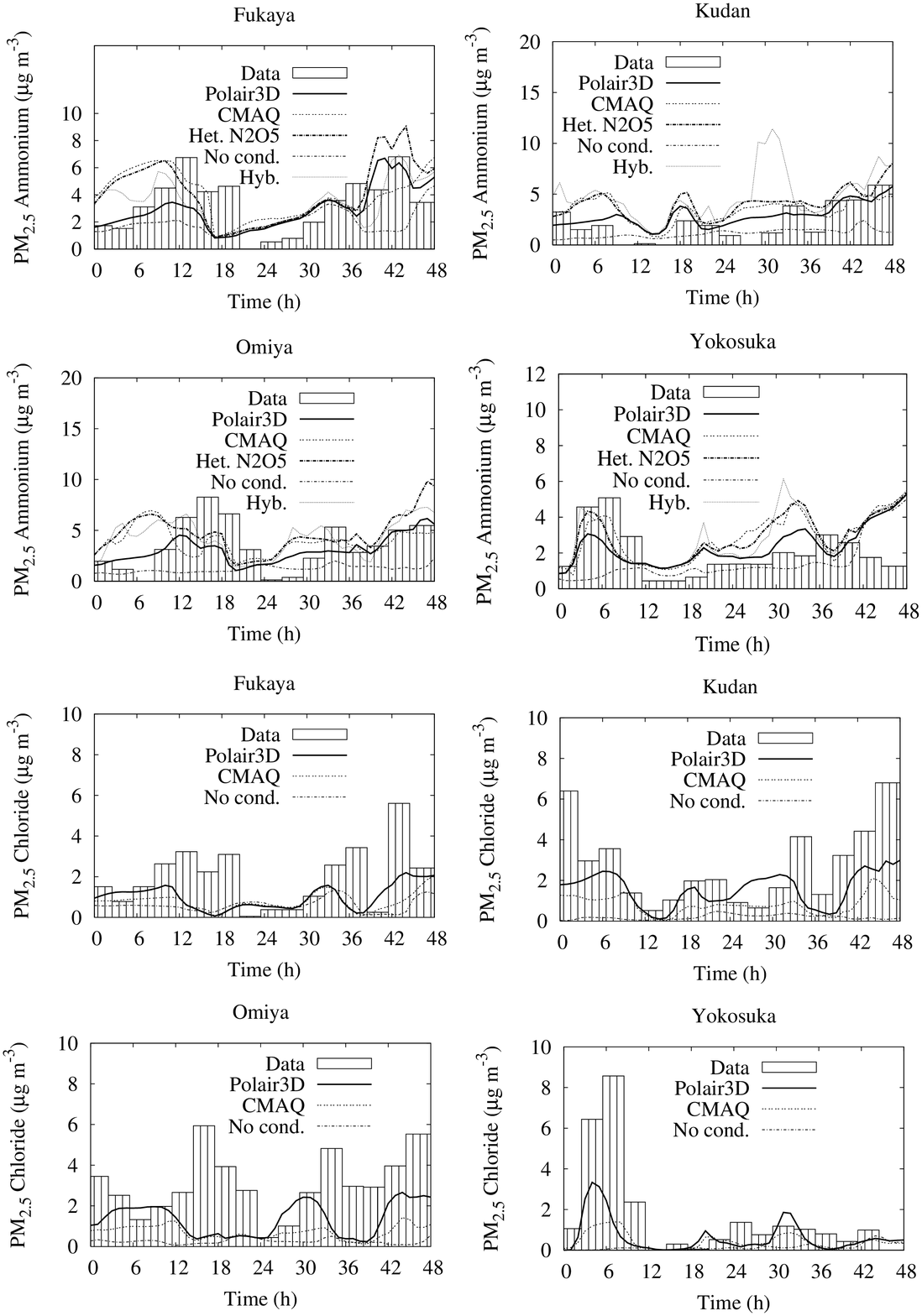}
        \end{center}
\caption{PM$_{2.5}$ concentration of ammonium and chloride at Fukaya, Yokosuka, Kudan, Omiya for 9 and 10 December 1999.}
\label{pm1999b}
\end{figure}

The presence of the meso-front in the afternoon of 9 December may be seen from the concentrations of $PM_{2.5}$ in Figures~\ref{pm1999} and~\ref{pm1999b}. Nitrate and ammonium concentrations are high in the afternoon before 6pm at Omiya and Fukaya, which are located in the northern part of the front where the pollutants accumulate. However, they are low at Yokosuka, which is located in the south part of the front. 
Kudan is located close to the front at 6pm. Concentrations are low at Kudan before 6pm and the concentrations of pollutants increase at Kudan from 6pm.
On 10 December, pollutants are observed to be high at the four stations, as a consequence of weak winds. 

Polair3D tends to over-estimate sulfate as shown by the $B_{NMBF}$, which is as high as $0.26$. As shown in Figures~\ref{pm1999} and~\ref{pm1999b}, sulfate is overestimated in the evening of the 10$^{th}$. For both ammonium and nitrate, although the position of the peaks in time is well predicted by Polair3D, they are under-estimated at Fukaya and Omiya on the 9$^{th}$. To find the reason of the discrepancies between simulations and measurements on the 9$^{th}$ for the peaks of nitrate and ammonium at Fukaya and Omiya, the sensitivity of the amplitude of these peaks to options in the aerosol module is studied in the next section.
Ammonium is overall over-estimated, while nitrate and chloride are under-estimated. However, for both nitrate and ammonium, the model has difficulties to predict the low concentrations in the night between the 9$^{th}$ and the 10$^{th}$ at Omiya and Fukaya, as confirmed by the scatter plots.
The reason for this discrepancy is also investigated in the next section.

\begin{table}
\caption{Correlation (corr), B$_{NMBF}$ and E$_{NMAEF}$ obtained with Polair3D for 9 and 10 December 1999.}
\begin{flushleft}
\begin{tabular}{cccc}
\tableline
         & corr    & B$_{NMBF}$     & E$_{NMAEF}$   \\
\tableline
Sulfate  & $66\%$ & $0.26$  & $0.33$   \\
Ammonium & $47\%$ & $0.05$  & $0.56$  \\
Nitrate  & $45\%$ & $-0.10$  & $0.58$   \\
Chloride & $36\%$ & $-1.15$  & $1.39$  \\
$SO_2$   & $44\%$ & $0.03$  & $0.56$ \\  
\tableline
\end{tabular}
\end{flushleft}
\label{RMSE1999}
\end{table}

\subsection{31$^{st}$ July and 1$^{st}$ August 2001}

\begin{figure}
	\begin{center}
	  \includegraphics[width=0.95\textwidth]{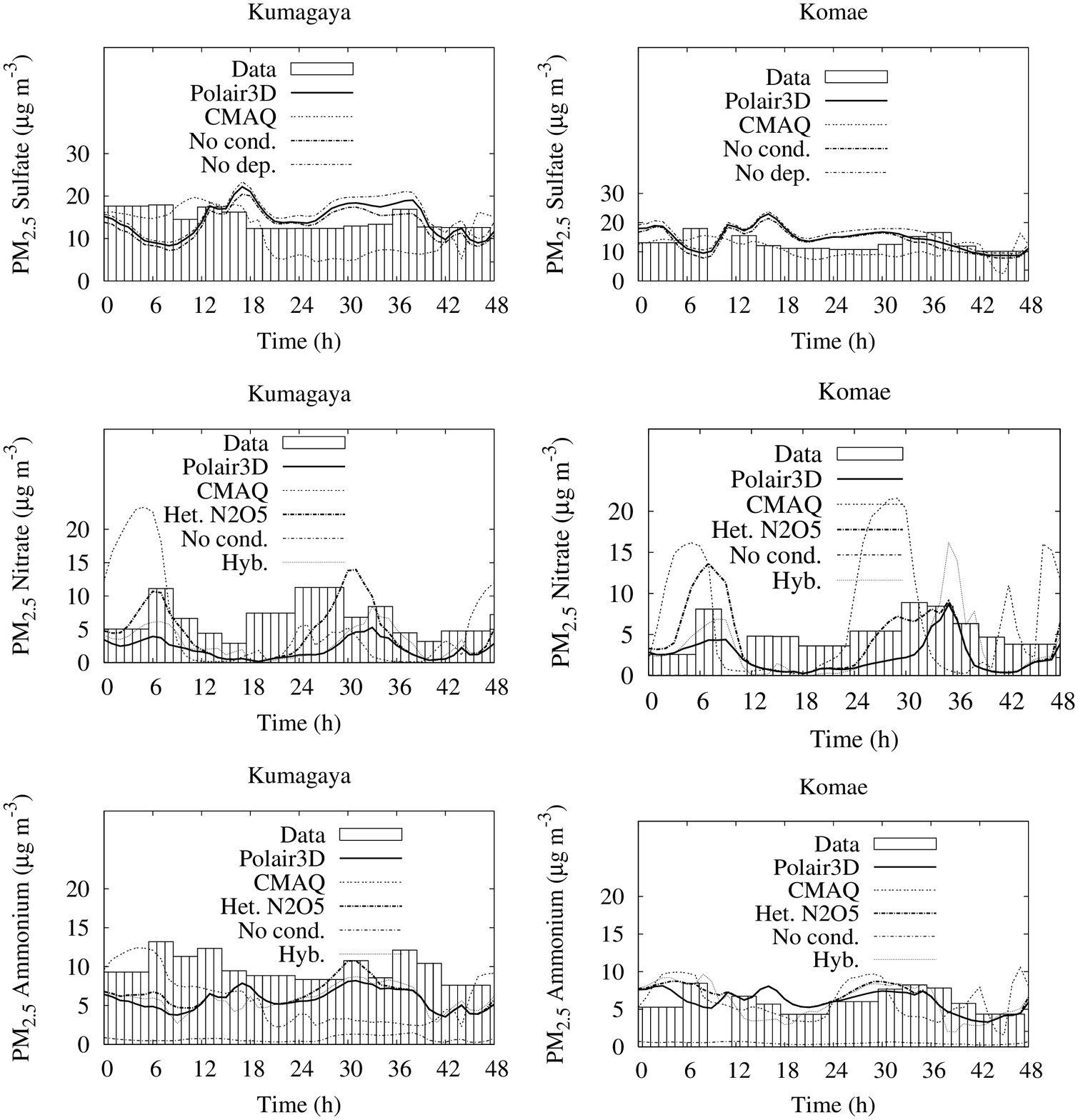}	  
        \end{center}
\caption{Comparisons of the $PM_{2.5}$ concentrations of sulfate, nitrate and ammonium at Kumagaya and Komae for 31 July and 1 August 2001 for different simulations.}
\label{pm2001}
\end{figure}

Comparisons of the results obtained with Polair3D are shown in Figure~\ref{pm2001} for PM$_{2.5}$. 

The statistical indicators are shown in Table~\ref{RMSE2001} for 31 July and 1 August 2001.  
As for 9 and 10 December 1999, the model satisfies the criteria suggested by \cite{yu-metrics} for sulfate. However, these criteria are again not satisfied for ammonium and nitrate.
Because chloride concentrations are very low in this summer run, they are not shown. 

As for the winter episode, the majority of hourly simulations falls within a factor 2 of the observations for sulfate (Figure~\ref{sp}). The majority of hourly simulation falls within a factor 2 of the observation at Komae, but scatter outside the reference lines is observed at Kumagaya. Considerable scatter is observed for nitrate at both Komae and Kumagaya with many simulated values falling outside the factor of two reference lines.

The correlation coefficient is low for sulfate (-$2\%$). For example, the model predicts a decrease in sulfate concentrations in the morning of 31 July at both Komae and Kumagaya (Figure~\ref{pm2001}), which is not confirmed by observations. 
Because this episode is a clear-sky one, cloud chemistry does not influence sulfate concentrations. Sulfate may be produced by the condensation of $H_2SO_4$, or it may be transported to the domain of study through boundary conditions. $H_2SO_4$ is either directly emitted or produced by the reaction of $SO_2$ with $OH$.
However, although the correlation coefficient is low for sulfate, good correlation is observed for $SO_2$ (Table~\ref{RMSE2001}). This suggests that uncertainty in sulfate concentrations is linked to uncertainty in the $H_2SO_4$ emissions and in the sulfate boundary conditions.

%For PM$_{2.5}$, the model performance criteria defined by \cite{boylan} are met by both models for inorganic components of PM$_{2.5}$, except for nitrate, for which a MFE as large as 100$\%$ is obtained. 

The sulfate concentrations, which are quite high all through the episode, are relatively well modeled. However, the measured concentrations do not vary much, whereas the model exhibits stronger time variations. 
For ammonium, although results are close to measurements at Komae, ammonium is under-estimated at Kumagaya. According to measurements, ammonium concentrations are twice as high in Kumagaya as in Komae. These high concentrations in Kumagaya are not reproduced by the models. Nitrate is severely under-estimated at both Kumagaya and Komae. 
The reasons for the discrepancies between measurements and modeled concentrations are investigated in the next section.

\begin{table}

\begin{flushleft}
\caption{Correlation (corr), B$_{NMBF}$ and E$_{NMAEF}$ obtained with Polair3D and CMAQ for 31 July and 1 August 2001.}
\begin{tabular}{cccc}
\tableline
         & corr    & B$_{NMBF}$     & E$_{NMAEF}$   \\
\tableline
Sulfate  & $-2\%$ & $0.11$  &  $0.32$ \\
Ammonium & $12\%$ & $-0.30$ & $0.49$ \\
Nitrate  &  $34\%$ & $-1.73$ & $1.85$ \\
$SO_2$   & $49\%$ & $0.04$ & $0.45$ \\
\tableline
\end{tabular}
\end{flushleft}
\label{RMSE2001}
\end{table}

\section{Impact of Aerosol Processes}

The impact of different processes on aerosol concentrations is investigated for both episodes.
The following processes are considered: nucleation, coagulation, condensation/evaporation, dry deposition, heterogeneous reactions. The impact of numerical schemes in the aerosol module is also studied: size-distribution and mode merging/splitting.

The impact of each of these physical and numerical processes is quantified by comparing a simulation S$_p$ where only a process $p$ is ignored (or its parameterization is changed) to the reference simulation S$_r$ where all the processes are taken into account. The simulations S$_p$ are successively: a simulation without condensation/evaporation, a hybrid simulation (thermodynamic equilibrium is only assumed for the smallest mode instead of being assumed for the four modes), a simulation with heterogeneous reactions (heterogeneous reactions are not taken into account in the reference simulation S$_r$), a simulation with only the N$_2$O$_5$ heterogeneous reaction, a simulation without the NO$_2$ heterogeneous reaction, a simulation without deposition, a simulation without nucleation, a simulation without coagulation, a simulation done using SIREAM instead of MAM (the log-normal size distribution is replaced by a sectional distribution), a simulation done using mode splitting instead of mode merging, and a simulation without neither mode splitting nor mode merging.
The quantification of the impact of each process $p$ is done by computing the B$_{NMBF}$ and the E$_{NMAEF}$ between S$_p$ and S$_r$ at the sites where measurements are made.
The larger in absolute value the B$_{NMBF}$ and the E$_{NMAEF}$ are, the stronger the impact of the process $p$ is. The statistics B$_{NMBF}$ and E$_{NMAEF}$ between CMAQ and $S_r$ are also computed in order to compare the impact of using a different chemistry transport model to the impact of each of the processes $p$. CMAQ takes into account condensation/evaporation with the same thermodynamic model as Polair3D (Isorropia), the heterogeneous reaction of N$_2$O$_5$, deposition, nucleation and mode merging. 
The comparisons of simulations done with Polair3D allow us to quantify the uncertainty related to the aerosol module. By using CMAQ, the simulated PM$_{2.5}$ concentrations may vary not only because of uncertainty in the aerosol module, but also because of uncertainties in chemistry, transport and diffusion. For example, even though Polair3D and CMAQ use the same meteorological fields, uncertainties in transport may be linked to the numerical schemes used, uncertainty in diffusion is related to the modeling of vertical diffusion.

Because the reference run does not always capture the observed behavior or because it may capture it for wrong reasons (e.g. inaccuracies in several processes cancel each other out), changing processes one at a time does not automatically give a reliable estimate of the relative importance of these processes in the real world, but only in the model. Similarily, the comparison of two CTMs does only give an estimate of the uncertainty related to transport, chemistry and diffusion, as these could be described inaccurately in both models. 

Tables~\ref{compmam1999} and~\ref{compmam2001} compare the reference simulation S$_r$ to different simulations $S_p$ for 9 and 10 December 1999 and 31 July and 1 August 2001 respectively.
For both episodes, the impact of heterogeneous reactions is large for nitrate and ammonium (the E$_{NMAEF}$s are as large as 1.95 and 0.79 respectively in winter). The HNO$_3$ produced by heterogeneous reactions condenses onto aerosols to form nitrate, and the available ammonia condenses to neutralize the nitrate. Concentrations of nitrate and ammonium increase by taking into account heterogeneous reactions, as shown by the positive B$_{NMBF}$. The impact of heterogeneous reactions on sulfate is small. The $H_2O_2$ concentration produced by heterogeneous reactions does not influence sulfate concentrations because aqueous chemistry is not taken into account.
The impact of condensation/evaporation is preponderant for ammonium, nitrate and chloride (for nitrate, the E$_{NMAEF}$ is as large as 2.44 in the winter episode and 96.4 in the summer episode). This impact is even larger than the impact of using CMAQ, or the impact of heterogeneous reactions. 
Condensation is largely preponderant over evaporation as seen by the negative B$_{NMBF}$.
Because of the low-volatility of sulfate, the impact of the hybrid scheme is small for sulfate (with a E$_{NMAEF}$ under 0.07). It is larger for other inorganic species, such as ammonium and nitrate, which are influenced by whether condensation/evaporation is computed dynamically or by assuming thermodynamic equilibrium.  The hybrid scheme mostly influences ammonium in the winter case with a E$_{NMAEF}$ of 0.41, and it influences mostly nitrate in the summer case with a E$_{NMAEF}$ of 0.49. 
Because coagulation and deposition do not differentiate the chemical composition of aerosols, their impacts are the same for each chemical species. The impact of deposition is between 0.08 and 0.10 for each species for both the winter and the summer simulations. Not taking deposition into account, leads as expected to an increase in aerosol concentrations as shown by the positive B$_{NMBF}$.
The impact of nucleation is small, with E$_{NMAEF}$ smaller than 0.06. Although a large number of nanometer particles are created by nucleation (the ternary nucleation scheme is several order of magnitudes larger than commonly
used binary nucleation schemes), the mass produced is small compared to the mass of PM$_{2.5}$. Furthermore, although the nucleated particles are made of sulfate and ammonium, nitrate and chloride are also influenced by nucleation, as they may condense onto freshly nucleated particles. 
Considering numerical processes, the impact of mode splitting and mode merging is negligible with E$_{NMAEF}$ smaller than 0.03. Mode merging/splitting influence the smaller particles, which do not contribute much to the mass of PM$_{2.5}$. 
The impact of using SIREAM, although not negligible, is not large for sulfate with  E$_{NMAEF}$ under 0.09. The impact is larger for nitrate, especially in the summer episode, where the E$_{NMAEF}$ reaches 0.22. A small impact on sulfate concentrations may correspond to larger impacts on other inorganic semi-volatile components. During condensation/evaporation, ammonium, nitrate and chloride are not only influenced by the size distribution of aerosols, but also by the internal composition because of their volatility. 
 
\subsection{9$^{th}$ and 10$^{th}$ December 1999}

In the winter episode of 1999, for sulfate, there is a dominant impact of both condensation and coagulation with E$_{NMAEF}$ as large as 0.27 and 0.21 respectively. These impacts are of the same order of magnitude as those of CMAQ (the E$_{NMAEF}$ is 0.23). For sulfate, the impacts of other processes are small, except for deposition, for which E$_{NMAEF}$ is equal to 0.10. Sulfate is produced by the condensation of H$_2$SO$_4$ and by the transport from outside the model domain through boundary conditions. Because sulfate decreases only by a factor 1.27 if condensation is not taken into account, the impact of long-range transport is likely to be high. The boundary condition for sulfate averaged over time, latitude and longitude is 3.0 $\mu$g~m$^{-3}$, against a mean concentration of 3.5 $\mu$g~m$^{-3}$ in the domain of study.
For other species, the impact of condensation/evaporation largely dominates, with
E$_{NMAEF}$ as large as 3.92 for chloride, 2.44 for nitrate and 1.04 for ammonium. 
This indicates that chloride, nitrate and ammonium are strongly influenced by local conditions. 
The impact of heterogeneous reactions is important for nitrate and ammonium with a E$_{NMAEF}$ as large as 1.95 and 0.79. The dominant heterogeneous reactions are the N$_2$O$_5$ and the NO$_2$ heterogeneous reactions. 
The E$_{NMAEF}$ is as high as 0.94 and 0.38 for nitrate and ammonium when only the N$_2$O$_5$ heterogeneous reaction is taken into account. A simulation without the NO$_2$ heterogeneous reaction but with all the other three gives similar results to the simulation with only the N$_2$O$_5$ heterogeneous reaction. 
The impact of the hybrid scheme is important for ammonium with a E$_{NMAEF}$ of 0.41, which is as large as the impact of using CMAQ (0.38). However, the impact is smaller for other chemical species, with a E$_{NMAEF}$ of 0.13 for nitrate against a E$_{NMAEF}$ of 1.0 when CMAQ is used.
Coagulation has an impact almost as large as the impact of condensation or the impact of CMAQ for sulfate with a E$_{NMAEF}$ of 0.21. Although the impact of coagulation is the same for each chemical species, it is less important for nitrate, ammonium and chloride in comparison to the impact of other processes.
Although nucleation is small, it is not negligible with a E$_{NMAEF}$ of 0.05 for sulfate and ammonium.

The impacts of the dominant processes during the episode are now investigated in more details, aiming at understanding the potential reasons for the discrepancies between measurements and simulated concentrations at the different stations (Figures~\ref{pm1999} and~\ref{pm1999b}).
As seen in section 5, sulfate is overall overpredicted. This is most of the time due to uncertainties in condensation and coagulation. For example,  
at Omiya and Kudan, Polair3D tends to overestimate the sulfate concentrations but CMAQ does not. Simulations without condensation or without coagulation give results that are similar to CMAQ, although both condensation and coagulation are taken into account in CMAQ.
As discussed in section 5, the peaks of nitrate and ammonium are underpredicted on the 9$^{th}$ at the stations above the meso-front: Omiya and Fukaya. When heterogeneous reactions are taken into account, the peaks are not underestimated any longer. However, they tend to be predicted earlier, stressing the difficulties of MM5 to accurately simulate the meso-front. Furthermore, the differences between Polair3D and CMAQ are often explained by the N$_2$O$_5$ heterogeneous reaction. If this heterogeneous reaction is taken into account, both Polair3D and CMAQ predict high nitrate and ammonium concentrations in the night and morning of the 9$^{th}$ at all four stations and in the night and in the morning of the 10$^{th}$ at Kudan, Omiya and Yokosuka. However, the high concentrations on the 10$^{th}$ are not observed, suggesting the need to revise the rate of the N$_2$O$_5$ heterogeneous reaction. For example, \cite{n2o5het} suggest a rate that vary with the aerosol composition, temperature and relative humidity. 
The low nitrate and ammonium concentrations between the 9$^{th}$ and the 10$^{th}$ at Omiya are overpredicted by CMAQ because of the heterogeneous reactions. However, although the low nitrate concentrations are quite well predicted by Polair3D when heterogeneous reactions are ignored, the low ammonium concentrations are still overpredicted. Although nitrate concentrations are very small all through the episodes if condensation is not taken into account, ammonium concentrations are larger than the lowest observed concentrations. The unability of the model to reproduce the low ammonium concentrations may therefore be linked to uncertainties outside the aerosol module, most likely in the boundary conditions.

\subsection{31$^{st}$ July and 1$^{st}$ August 2001}

In the summer episode, for sulfate, the impacts of the different processes are small compared to the impact of using CMAQ. The E$_{NMAEF}$ for the simulation with CMAQ is equal to 0.55 against 0.08 for other simulations such as the one without condensation, the one using SIREAM and the one without deposition. The impact of condensation on sulfate is not as strong as it is in the winter episode. This suggests that sulfate mostly comes from long-range transport. 
Whereas in the winter case, the averaged sulfate in the domain is larger than the averaged sulfate used for boundary conditions, the opposite holds in the summer case. The boundary condition for sulfate averaged over time, latitude and longitude is 16.6 $\mu$g~m$^{-3}$, against a mean concentration of 13.5 $\mu$g~m$^{-3}$ in the domain of study.
In fact, Mount Oyama of Miyake Island, which is located about 180km south of central Tokyo, was in eruption at that time (\cite{miyakejima}). Sulfate concentrations may be more sensitive to transport or diffusion processes, such as the parameterization used to model the vertical diffusion (\cite{mallet05uncertainty})
However, condensation influences to a great extent the nitrate and ammonium concentrations, with a E$_{NMAEF}$ as large as 96.4 and 9.4 respectively. The impact of condensation on ammonium and nitrate is large compared to the impact of using CMAQ. Ammonium and nitrate are produced locally.
Although condensation is largely a dominant process for the ammonium and nitrate concentrations, the influence of the hybrid simulation, where the thermodynamic equilibrium assumption is removed, is limited with a E$_{NMAEF}$ of only 0.49 for nitrate and 0.18 for ammonium. 
The relative importance of the hybrid scheme versus condensation is higher in the winter episode than in the summer episode.
Perhaps, this is because thermodynamic equilibrium is reached quickly under high temperatures (the mean temperature is 295K in the summer episode and 278K in the winter episode) and high pollutant concentrations (the averaged PM$_{2.5}$ concentration is 20.1$\mu$g~m$^{-3}$ in the summer episode and 13.3$\mu$g~m$^{-3}$ in the winter episode) (\cite{ed_thermo_WS}).
Nitrate concentrations, and ammonium concentrations to a lesser extent, are also strongly influenced by heterogeneous reactions with a E$_{NMAEF}$ of 1.20 and 0.12 respectively. As for the winter case, the dominant heterogeneous reactions are the N$_2$O$_5$ and the NO$_2$ heterogeneous reactions. The E$_{NMAEF}$ for the N$_2$O$_5$ heterogeneous reactions is as large as 0.95 for nitrate and 0.09 for ammonium.
The impact of using SIREAM is not small although not preponderant: it is as large as the impact of deposition and condensation for sulfate, and it is as large as the impact of the hybrid scheme for ammonium and nitrate.
Mode splitting, mode merging, nucleation and coagulation are negligible during this episode. 
Nucleation and coagulation are negligible in the summer case but not in the winter case. The mean temperature and relative humidity in the winter case are 278K and 49$\%$, while they are 295K and 76$\%$ in the summer case. According to \cite{korhonen}, under these conditions of temperature and relative humidity, the nucleation rate is about 6 times higher in the winter case than in the summer case.
The impact of coagulation is larger than the impact of nucleation, because coagulation does not only influence particles when they are freshly nucleated, but coagulation also influences these particles as they grow by condensation.

\begin{figure}
	\begin{center}
	  \includegraphics[width=0.95\textwidth]{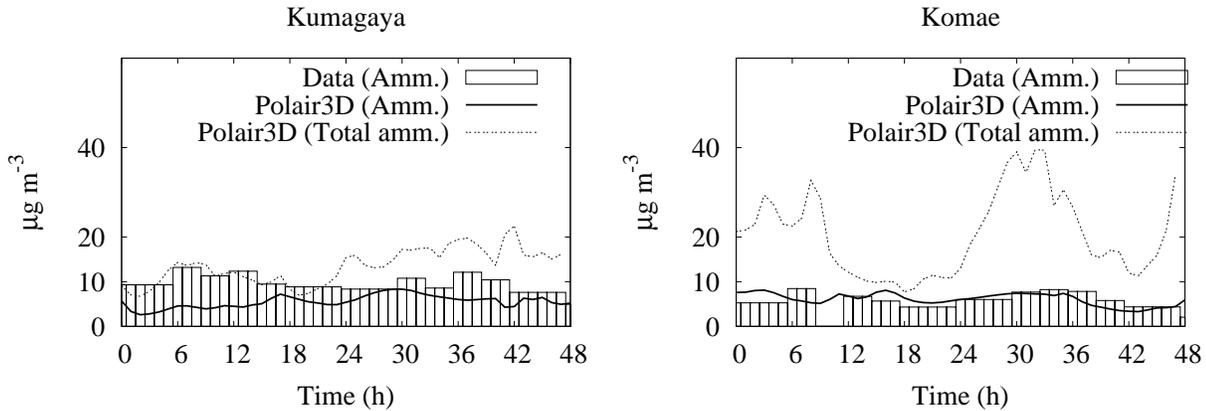}
        \end{center}
\caption{Concentrations of ammonium (modeled and observed) and total ammonium at Kumagaya and Komae for 31 July and 1 August 2001.}
\label{totamm}
\end{figure}

\begin{figure}
	\begin{center}
	  \includegraphics[width=0.95\textwidth]{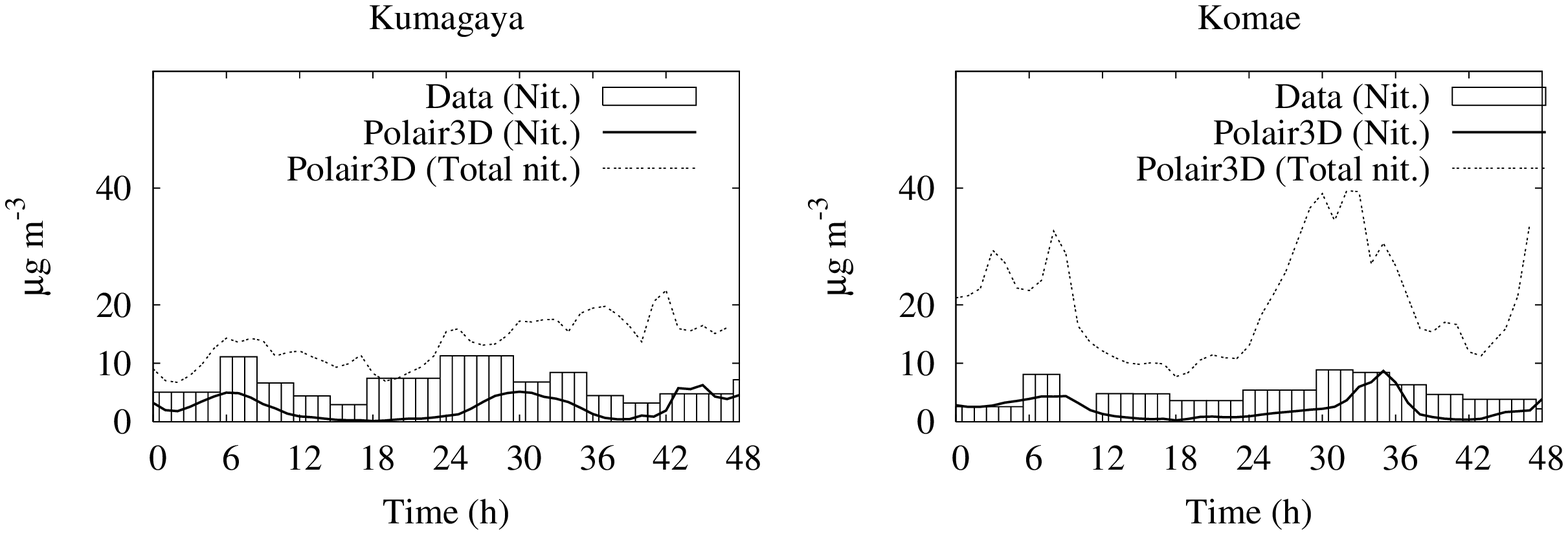}
        \end{center}
\caption{Concentrations of nitrate (modeled and observed) and total nitrate at Kumagaya and Komae for 31 July and 1 August 2001.}
\label{totnit}
\end{figure}

The impacts of the dominant processes during the episode are now investigated in more details, aiming at understanding  the potential reasons for the discrepancies between measurements and simulated concentrations at the different stations (Figure~\ref{pm2001}).
The sulfate concentrations are very little sensitive to the options used in the aerosol module in Polair3D, but large differences are observed between CMAQ and Polair3D. For example, at Kumagaya, Polair3D predicts lower sulfate concentrations than CMAQ in the night and morning of the 31$^{st}$, whereas it predicts higher sulfate concentrations later in the episode. Furthermore, the models exhibit stronger diurnal variations compared to observations.
Because sulfate concentrations mostly come from long-range transport, and are little sensitive to options used in the aerosol module, the uncertainties in the sulfate concentrations are linked to uncertainties in the meteorology and boundary conditions. Polair3D and CMAQ use the same meteorological fields, except for the vertical diffusion, which is computed following \cite{troen} in Polair3D and following \cite{cmaq} (K-theory) in CMAQ. Uncertainty due to vertical diffusion is likely to be important as stressed by \cite{mallet05uncertainty}. 
Boundary conditions are provided on the vertical grid of CMAQ, which is used for the continental run, and they are projected onto the vertical grid of Polair3D. Uncertainties due to boundary conditions may not only be linked to this projection, but also to the projection of moments. Because the boundary conditions are obtained from CMAQ, they are given as the moments of order zero, two and three for particulate matter. However, a projection is required by Polair3D which uses the moments of order zero, three and six to represent particulate matter.
Ammonium and nitrate are strongly influenced by condensation. Their concentrations are very low, and almost zero for nitrate, if condensation is not taken into account. 
As ammonia may condense onto aerosols to neutralize the sulfate, ammonium follows the diurnal evolution predicted for sulfate (the correlation between computed sulfate and ammonium is as high as 87$\%$). Errors on ammonium concentrations are partly due to errors on sulfate concentrations. Another cause of errors may come from uncertainties in total ammonium. Ammonium concentrations are better predicted at Komae than at Kumagaya. As shown in Figure~\ref{totamm}, the total ammonium concentration computed by Polair3D is higher at Komae than at Kumagaya, even though ammonium concentration in the aerosol phase is higher at Kumagaya. In the first part of the episode, at Kumagaya, the total ammonium computed is as low as the measured ammonium in the aerosol phase. Even though ammonium is systematically underestimated at Kumagaya, some discrepancies between modeled and observed ammonium can be explained by processes in the aerosol module at Komae. For example, whereas Polair3D overestimates ammonium concentrations in the afternoon of the 31$^{st}$, the ammonium concentrations are underestimated if the hybrid scheme is used.
Nitrate is strongly underpredicted by Polair3D. 
However, the peaks of nitrate at night are better predicted when the N$_2$O$_5$ heterogeneous reaction is taken into account, although these peaks are sometimes overpredicted by CMAQ. As for total ammonium, total nitrate concentration is much higher at Komae than at Kumagaya (Figure~\ref{totnit}). However, even though the total nitrate concentration is high at Komae, 
the high concentrations of nitrate in the aerosol phase in the afternoon of the 31$^{st}$ at Komae are not reproduced even if the hybrid scheme is used. 
\cite{yu-nitrate} suggest that measurement uncertainties in sulfate and total ammonium, i.e. ammonium plus ammonia, can account for most of the discrepancies between the model predictions and observations in partitioning of aerosol nitrate.

\begin{table}
\caption{Comparison of Polair3D-MAM to different simulations for 9 and 10 December 1999: B$_{NMBF}$ (B) and E$_{NMAEF}$ (E).}
\begin{flushleft}
\begin{tabular}{ccccccccc}
\tableline
              & \multicolumn{2}{c}{Sulfate}     & \multicolumn{2}{c}{Ammonium}     & \multicolumn{2}{c}{Nitrate}  & \multicolumn{2}{c}{Chloride}  \\
\tableline
                & B & E & B & E & B & E & B & E \\
\tableline
No condensation & $-0.27$ & $0.27$ & $-1.03$ & $1.04$  & $-2.44$ & $2.44$ & $-3.82$ & $3.92$ \\
Hybrid       & $0.07$ & $0.07$    & $0.37$ & $0.41$ & $0.12$ & $0.13$ & $0.02$ & $0.07$ \\
Het. React.  & $0.03$ & $0.03$    & $0.79$ & $0.79$ & $1.95$ & $1.95$ & $0.01$ & $0.08$ \\
Only N$_2$O$_5$ het. react.  & $0.01$ & $0.01$    & $0.38$ & $0.38$ & $0.94$ & $0.94$ & $0.03$ & $0.04$ \\
No NO$_2$ het. react.  & $0.01$ & $0.01$    & $0.38$ & $0.38$ & $0.93$ & $0.93$ & $0.03$ & $0.04$ \\
No deposition & $0.10$ & $0.10$    & $0.09$ & $0.09$   & $0.10$ & $0.10$   & $0.09$ & $0.09$  \\
No nucleation & $0.05$ & $0.05$    & $0.05$ & $0.05$   & $0.06$ & $0.06$   & $0.06$ & $0.06$ \\
No coagulation & $-0.21$ & $0.21$ & $-0.22$ & $0.22$ & $-0.25$ & $0.25$   & $-0.18$ & $0.18$ \\
SIREAM        & $0.03$ & $0.06$    & $-0$ & $0.05$  & $-0.09$ & $0.09$ & $-0.08$ & $0.08$ \\
Splitting     & $0.01$ & $0.02$    & $0.01$ & $0.02$   & $0.01$ & $0.02$   & $0.02$ & $0.02$ \\
No merging - no splitting   &  $-0.01$ & $0.03$  & $-0.01$ & $0.03$   & $-0.01$ & $0.03$ & $0$ & $0.02$ \\
CMAQ          & $-0.21$ & $0.24$  & $0.26$ & $0.38$ & $0.85$ & $1.00$ & $-0.63$ & $0.69$ \\
\tableline
\end{tabular}
\end{flushleft}
\label{compmam1999}
\end{table}

\begin{table}
\caption{Comparison of Polair3D-MAM to different simulations for 31 July and 1 August 2001: B$_{NMBF}$ (B) and E$_{NMAEF}$ (E).}
\begin{flushleft}
\begin{tabular}{ccccccc}
\tableline
              & \multicolumn{2}{c}{Sulfate}     & \multicolumn{2}{c}{Ammonium}     & \multicolumn{2}{c}{Nitrate}  \\
\tableline
                & B & E & B & E & B & E \\
\tableline
No condensation & $-0.08$ & $0.08$ & $-9.40$ & $9.40$  & $-96.4$ & $96.4$ \\
Hybrid       & $0.06$ & $0.06$     & $0$ & $0.18$   & $0.40$ & $0.49$ \\
Het. React. & $0$ & $0$    & $0.12$ & $0.12$   & $1.20$ & $1.20$ \\
Only N$_2$O$_5$ het. react.  & $0$ & $0$    & $0.09$ & $0.09$ & $0.95$ & $0.95$ \\
No NO$_2$ het. react.  & $0$ & $0$    & $0.09$ & $0.09$ & $0.92$ & $0.93$ \\
No deposition & $0.08$ & $0.08$    & $0.08$ & $0.08$     & $0.08$ & $0.08$ \\
No nucleation & $0.01$ & $0.01$    & $0.01$ & $0.01$     & $0.02$ & $0.02$ \\
No coagulation & $-0.01$ & $0.01$  & $-0.01$ & $0.01$    & $-0.02$ & $0.02$  \\
SIREAM        & $-0.08$ & $0.08$   & $-0.11$ & $0.11$  & $-0.22$ & $0.22$ \\
Splitting     & $0.02$ & $0.02$    & $0.02$ & $0.02$     & $0.02$ & $0.02$ \\
No merging - no splitting & $0.02$ & $0.02$    &  $0.02$ & $0.02$  & $0.02$ & $0.02$  \\
CMAQ          & $-0.32$ & $0.55$ & $-0.01$ & $0.48$  & $1.87$ & $2.51$ \\
\tableline
\end{tabular}
\end{flushleft}
\label{compmam2001}
\end{table}

\section{Conclusion}

In this article, two high-pollution episodes over Greater Tokyo are studied: one during the winter 1999 (9 and 10 December) and one during the summer 2001 (31 July and 1 August 2001). For each of these episodes, the chemistry transport model Polair3D is compared to measurements for inorganic components of PM$_{2.5}$.

For sulfate, error statistics are in agreement with model performance criteria (\cite{yu-metrics}). Inorganic components of PM$_{2.5}$ remain overall well modeled except for nitrate in the summer episode.

To understand to which extent the aerosol processes modeled in Polair3D influence the particle concentrations during the summer and the winter episodes, different simulations are made where only one process differs from the default or reference simulation. The following physical/chemical processes are considered: nucleation, coagulation, condensation/evaporation, whether condensation is modeled dynamically or using the thermodynamic equilibrium assumption, dry deposition, heterogeneous reactions. For numerical processes, the impact of mode merging/mode splitting and the impact of the size distribution (modal versus sectional) are evaluated. A comparison of the impact of each aerosol process described above to the impact of using the CTM CMAQ allows us to assess the importance of using different parameterizations and numerical schemes not only for aerosol processes but also for chemistry, transport and diffusion.

%\vskip 2pt
%For both the winter and the summer cases, the impact of mode splitting and mode merging is negligible. The impact of dry deposition and the impact of heterogeneous reactions on sulfate are small although not negligible.
%However, the impact of heterogeneous reactions is large for species other than sulfate such as nitrate. The impact of condensation/evaporation is preponderant for ammonium, nitrate and chloride. The impact of the hybrid scheme is small for sulfate but it may be large for other inorganic species.  The hybrid scheme influences mostly ammonium in the winter case, and it influences mostly nitrate in the summer case. The impact of using SIREAM, although not negligible, is not large for sulfate, but the impact is larger for nitrate, especially in summer. Although coagulation has a large impact in winter, it is negligible in summer.

This study illustrates that the impact of aerosol processes on aerosol concentrations differs depending on local conditions and aerosol chemical components. For example, in the summer episode, for sulfate, the impact of long-range transport largely dominates. In the winter episode, sulfate is mostly impacted by condensation, coagulation, long-range transport, and deposition to a lesser extent. Whereas nucleation and coagulation are negligible in the summer episode, they are not in the winter episode. 
The impact of coagulation is larger in the winter episode than in the summer episode, because the number of small particles is higher in the winter episode as a consequence of nucleation. 
The impact of condensation/evaporation is dominant for ammonium, nitrate and chloride in both episodes. However, the impact of the thermodynamic equilibrium assumption is limited. The impact of heterogeneous reactions is large for nitrate and ammonium. The dominant heterogeneous reactions are the NO$_2$ $\rightarrow$ 0.5 HONO + HNO$_3$ and the N$_2$O$_5$ $\rightarrow$ 2 HNO$_3$ reactions. 
The impact of using a sectional representation of the size distribution is not negligible, and it is higher for ammonium and nitrate than for sulfate.
The impact of mode merging/mode splitting is negligible in both episodes.

The comparison of the different runs also allows us to understand discrepancies between observed and simulated inorganic PM$_{2.5}$ at different stations. Heterogeneous reactions appear to be crucial in predicting the peaks of nitrate and ammonium in the winter episode. However, heterogeneous reactions sometimes lead to concentrations that are too high, suggesting the need for a more detailed parameterisation of the reaction rates.

Although the impact of mode merging and mode splitting is negligible on PM$_{2.5}$  concentrations, it may influence the size distribution of aerosols.
Larger differences between the different runs may be observed by comparing the size distribution of aerosols or the concentrations of smaller particles such as PM$_{1}$. In particular, the impact of using a sectional rather than a modal model would be larger on the size distribution than on the mass of PM$_{2.5}$. The impact of nucleation and whether condensation is computed dynamically or using the thermodynamic assumption would be larger on PM$_{1}$ than on PM$_{2.5}$.

\begin{acknowledgments}

This work was supported by the Canon Foundation in Europe, and by the program Primequal under the project PAM (Multiphase Air Pollution).
We would like to thank the anonymous reviewers for their constructive comments on the manuscript.

\end{acknowledgments}

\appendix
\section*{Appendix A: Statistical Indicators}

The following indicators are computed in order to
evaluate error statistics for model-to-data comparisons.  Let $\left( o_i
\right)_i$ and $\left( c_i \right)_i$ be the observed and the modeled
concentrations, where $i$ is over $n$ time series and locations, and $\bar{o} = \sum_{i=1}^n o_i$ and $\bar{c} = \sum_{i=1}^n c_i$ the averaged observed and modeled concentrations respectively

\begin{itemize}

\item Correlation
\begin{equation}
  \frac{\sum_{i=1}^n \left( c_i - \bar{c} \right) \left( o_i - \bar{o} \right) }{\sqrt{\sum_{i=1}^n \left( c_i - \bar{c} \right)^2} \sqrt{\sum_{i=1}^n \left( o_i - \bar{o} \right)^2}}
\end{equation}
\begin{equation}
  \mbox{with: } \bar{o} = \frac{1}{n} \sum_{i=1}^n o_i \; \mbox{and} \; \bar{c} = \frac{1}{n} \sum_{i=1}^n c_i 
\end{equation}

\item Normalized mean bias factor (B$_{NMBF}$)
\begin{equation}
\frac{\bar{c} }{ \bar{o} } - 1 \; \mbox{if } \bar{c} \ge \bar{o} 
\end{equation}
\begin{equation}
1 - \frac{\bar{o}  }{ \bar{c}} \; \mbox{if } \bar{c} < \bar{o} 
\end{equation}

\item Normalized mean absolute error factor (E$_{NMAEF}$)
\begin{equation}
\frac{\sum_{i=1}^n \left| c_i - o_i \right|}{\bar{o}} \; \mbox{if } \bar{c} \ge \bar{o} 
\end{equation}
\begin{equation}
\frac{\sum_{i=1}^n \left| c_i - o_i \right|}{\bar{c}} \; \mbox{if } \bar{c} < \bar{o} 
\end{equation}

\end{itemize}

%% ------------------------------------------------------------------------ %%
%
%  TEXT
%
%% ------------------------------------------------------------------------ %%
%
%  REFERENCE LIST AND TEXT CITATIONS
%
%% ------------------------------------------------------------------------ %%
%
% If you use BiBTeX for your References, please do not send
% your bibliography database. Copy the reference list
% from your .bbl file into your article file before submission:
%
%1. Run LaTeX on your LaTeX file.
%
%2. Run BiBTeX on your LaTeX file.
%
%3. Open the new .bbl file containing the reference list and
%copy all the contents into your LaTeX file after the
%acknowledgments section;
%
%4. Comment out the old \bibliographystyle and \bibliography commands.
%
%5. Run LaTeX on your new file before submitting.

%Failure to follow these instructions will require manual
%intervention through hard keying of information,
%which can introduce errors.

\end{article}

\end{document}